\documentclass[prd,amsmath,amssymb,floatfix]{revtex4}

\usepackage{graphicx}

\begin{document}

\title{Biological Homochirality and the Search for Extraterrestrial Biosignatures}

\author{Marcelo Gleiser}
\email{mgleiser@dartmouth.edu}

\affiliation{Department of Physics and Astronomy, Dartmouth College
Hanover, NH 03755, USA}

\date{March 18, 2022}

\begin{abstract}
\centerline{\bf ABSTRACT}
\vspace{0.2 in}
Most amino acids and sugars molecules occur in mirror, or chiral, images of each other, knowns as enantiomers. However, life on Earth is mostly homochiral: proteins contain almost exclusively L-amino acids, while only D-sugars appear in RNA and DNA. The mechanism behind this fundamental asymmetry of life remains unknown, despite much progress in the theoretical and experimental understanding of homochirality in the past decades. We review three potential mechanisms for the emergence of biological homochirality on primal Earth and explore their implications for astrobiology: the first, that biological homochirality is a stochastic process driven by local environmental fluctuations; the second, that it is driven by circularly-polarized ultraviolet radiation in star-forming regions; and the third, that it is driven by parity violation at the elementary particle level. We argue that each of these mechanisms leads to different observational consequences for the existence of enantiomeric excesses in our solar system and in exoplanets, pointing to the possibility that the search for life elsewhere will help elucidate the origins of homochirality on Earth.

\end{abstract}

\maketitle


\section{Introduction: A Bit of History}

In 1815, the French physicist and chemist Jean-Baptiste Biot discovered that when light travelled through liquid solutions made out of a number of naturally occurring organic products, its polarization was affected. Pasteur was well aware of Biot's studies \cite{Gleiser_book}. As he wrote in a set of lecture notes from 1860, ``[Biot] quite definitely concluded that the action produced by the organic bodies was a molecular one, peculiar to their ultimate particles and depending on their individual constitution.'' \cite{Pasteur} The ``action'' Pasteur referred to was the ability of these natural organic compounds to rotate the polarization direction of light. With remarkable prescience, Biot had conjectured that such property was related to something going on at the molecular level. Pasteur put Biot's conjecture into firm ground, showing that the optical properties of certain organic compounds--the way they interacted with light--resulted from the spatial structure of their individual molecules. Building upon Biot's research, Pasteur established that when linearly polarized light passed through a solution of tartaric acid synthesized in the lab, nothing happened: the synthetic solution was optically inactive. But when polarized light passed through a solution containing acid extracted from grapes, and thus from a living entity, its polarization direction changed.

Pasteur realized that since both substances had identical chemical properties, their molecules had the same types of atoms. What then could cause such puzzling asymmetric behavior? Could living and nonliving substances, even if apparently identical, have different properties? He examined the crystals from both substances under a microscope. He noted that whereas the lab-synthesized acid had two kinds of crystals, the acid from grapes had only one. With tremendous patience, he separated samples of both crystals using tweezers. Passing light through two solutions made with each of them, he demonstrated that the different crystals rotated the polarization plane of light in opposite directions:
``I carefully separated the crystals which were [asymmetric] to the right from those [asymmetric] to the left, and examined their solutions separately in the polarizing apparatus. I then saw with no less surprise than pleasure that the crystals [asymmetric] to the right deviated the plane of polarization to the right, and that those [asymmetric] to the left deviated it to the left; and when I took an equal weight of each of the two kinds of crystals, the mixed solution was indifferent towards the light in consequence of the neutralization of the two equal and opposite individual deviations.'' \cite{Pasteur}

Pasteur's remarkable finding was that the naturally-occurring compound only appears in one of its two possible forms while the synthetic one appears in both. Was life selecting a specific molecular orientation?
Continuing with his investigation, Pasteur showed that many organic compounds extracted from living organisms had the same biased optical properties. In one experiment, he added mold to a synthetic sample of tartaric acid. Initially, there was no optical activity, as expected. But as the mold grew, so did the optical activity of the sample. Furthermore, the increasing rotation was in the same direction of the naturally occurring acid. There was only one possible conclusion: life had a molecular bias. As Pasteur later wrote, ``The Universe is dissymmetric and I am persuaded that life, as it is known to us, is a direct result of the asymmetry of the Universe or of its indirect consequences.''

Think of proteins as long chains of amino acids, pearl necklaces where each pearl is a molecular building block. Imagine that a left-handed (levorotatory) amino acid is a white pearl and a right-handed (dextrorotatory) amino acid a black pearl. Life has a clear (but not exclusive) preference for white pearl necklaces: the crucial molecules for life, proteins, are built from asymmetric backbones. The same is true for the sugar backbones of RNA and DNA. However, in this case the bias goes the opposite way: the sugars are dextrorotatory. It is hard to avoid the suspicion that this molecular bias is somehow related to the origin of life itself. Pasteur was the first to speculate as such: 
``Why even right or left substances at all? Why not simply non-asymmetric substances; substances of the order of inorganic nature? There are evidently causes for these curious manifestations of the play of molecular forces\dots. Is it not necessary and sufficient to admit that at the moment of the elaboration of the primary principles in the vegetable organism, an asymmetric force is present?'' \cite{Pasteur}

Pasteur had essentially posed the question we are still asking, whether homochirality has a specific biological function that, somehow, is related to some yet unknown fundamental causal mechanism. Although books \cite{Janoschek,Wagniere, Hochberg21, Cline}, reviews \cite{Blackmond2019,Martin2007}, and other contributions to this volume will no doubt examine different potential mechanisms to promote and amplify a small initial chiral bias of biomolecular backbones, in this work I propose to examine a related question that is deeply correlated to the issue of causation, to wit, the range of biological homochirality across astronomical distances and what it can teach us about the mechanism(s) driving homochirality. In brief, whether the enantiomeric excess we observe on Earth is relegated exclusively to our planet; whether the same enantiomeric excess operates on our solar system and possibly others in our local neighborhood; ad finally, whether the same bias toward homochirality exists across the Universe. Each of these three possibilities is a consequence of a different abiotic causal mechanism driving an initial chiral bias. Looking for enantiomeric excesses in planets and moons of our solar system and elsewhere in the Universe will have much to teach us about the nature of homochirality on Earth and elsewhere in the cosmos. We note that although it is often stated that life on Earth is homochiral, and that amino acids are exclusively  levorotatory and sugars dextrorotatory, this is absolutely not the case. There are plenty of D-amino acids, such as D-alanine, and L-sugars, like L-arabinose, that play an essential role in our biosphere. For a review see Ref. \cite{Finefield2012}. Thus, when we use the term ``homochiral,'' we are referring to a preponderant excess of one enantiomer over the other, not an absolute dominance.

This paper is organized as follows: In Section II ({\it Mechanisms for Chiral Bias}), I review three possible causes for biochirality. Namely, random environmental processes on primal Earth \cite{GTW}; circularly polarized UV light in stellar-forming regions \cite{Bailey,CPL2,Lucas05}, and parity-violating weak neutral currents at the elementary-particle level \cite{WNC0,Salam91,KN85, WNC3, WNC4}. In Section III ({\it Astronomical Impact of Different Chiral Biasing Mechanisms}), I review how each of these mechanisms impacts different astronomical ranges. In Section IV ({\it Concluding Remarks: Observing Biological Chirality in the Universe}), I discuss how such mechanisms could be differentiated through observations, linking stereochemistry and astrobiology. I conclude by summarizing the discussion and describing possible avenues for {\it in situ} and remote sensing in present and future research.

\section{Mechanisms for Chiral Bias}

This section reviews three potential abiotic mechanisms for promoting chiral bias at the molecular level. Each will have a specific range of astronomical impact and thus be of relevance for astrobiological research and future searches for organic materials in our solar system and, via remote observation, in star-forming regions and exoplanets that display promising biosignatures. The author apologizes beforehand for the unevenness of the discussion, weighted disproportionately toward some of his work. However, with the added references and discussions, the interested reader can certainly pursue further details.

\subsection{Punctuated Chirality: Planetary Bias}

\noindent In reference \cite{GTW}, Gleiser, Thorarinson, and Walker (GTW) proposed a mechanism dubbed ``Punctuated Chirality'' whereby the drive toward homochirality on Earth was the product of random environmentally-driven fluctuations that critically affected the prebiotic, molecular-forming mix of organic compounds, possibly many times over. The starting point was Sandars' polymerization model \cite{Sandars03}, a generalization of Frank's pioneering approach \cite{Frank53}, featuring autocatalysis with enantiomeric cross-inhibition. Consider a left-handed polymer $L_n$, made of $n$ left-handed monomers, $L_1$. It may grow by adding another left-handed monomer with a rate $k_s$, or be inhibited by adding a right-handed monomer $D_1$ with a rate $k_I$. (Note that we denote D-compounds by the letter ``D'' as opposed to the notation set in Sandars' work where such molecules were denoted as ``R''.) The reaction network for $n = 1, \ldots , N$, where $N$ is the maximum polymer length in the system, can be written as:

\begin{eqnarray}\label{rxnnetwork}
L_n + L_1 & \stackrel{2k_S}{\rightarrow} L_{n+1}, \nonumber \\
L_n + D_1 & \stackrel{2k_I}{\rightarrow} L_nD_1, \nonumber \\
L_1 + L_nD_1 & \stackrel{k_S}{\rightarrow} L_{n+1}D_1, \nonumber \\
D_1 + L_nD_1 & \stackrel{k_I}{\rightarrow} D_1L_nD_1, \nonumber \\
\end{eqnarray} 
supplemented by reactions for $D$-polymers by interchanging $L \rightleftharpoons D$,  and by the production rate of monomers from the substrate: $S \stackrel{k_C C_L}{\longrightarrow} L_1$;  $S \stackrel{k_C C_D}{\longrightarrow} D_1$. $C_{L (D)}$ determine the enzymatic enhancement of $L (D)$-handed monomers, usually assumed to depend on the largest polymer in the reactor pool, $C_{L(D)} = L_N(D_N)$ \cite{Sandars03}, or on a sum of all polymers \cite{WC}. Soai's group obtained the best-known illustration of this autocatalytic mechanism with enantiomeric cross-inhibition \cite{Soai}, with dimers ($N = 2$) as catalysts \cite{Blackmond04}.

A set of coupled, nonlinear ordinary differential equations for the various concentrations, $[L_1]$,  $[D_1]$, \ldots, $[L_n]$, $[D_n]$, describes the time evolution of the reaction network of eq. \ref{rxnnetwork}, supplemented by the equation for the substrate, $d[S]/dt = Q (Q_L + Q_D)$, where $Q$ is the substrate's production rate, and $Q_L - Q_D = k_Cf[S](C_L - C_D)$ gives the net chiral excess in monomer production. $f$ is the enzymatic fidelity, usually set to unity to maximize chiral separation. GTW showed that starting as racemates, numerical solutions for polymerization reactions with $N = 2, 5$, and $\infty$ evolve toward homochirality. This is also the case for $N = 2$ within the adiabatic approximation, where the rate of change for dimers and the substrate is assumed to be much slower than that of monomers, that is, when $k_{S,I} << k_C$ \cite{BM}. In Gleiser and Walker \cite{GW}, a detailed study of the polymerization reaction network for various values of $N$ has shown that the trends remain true when the effects of spatial dynamics are considered. Gleiser and Walker also concluded that although the adiabatic approximation predicts faster approach to steady-state conditions when compared with the full $N = 2$ model, it does produce the correct asymptotic values for the various concentrations.

\begin{figure}
\centerline{\includegraphics[width=4.5in,height=3.25in]{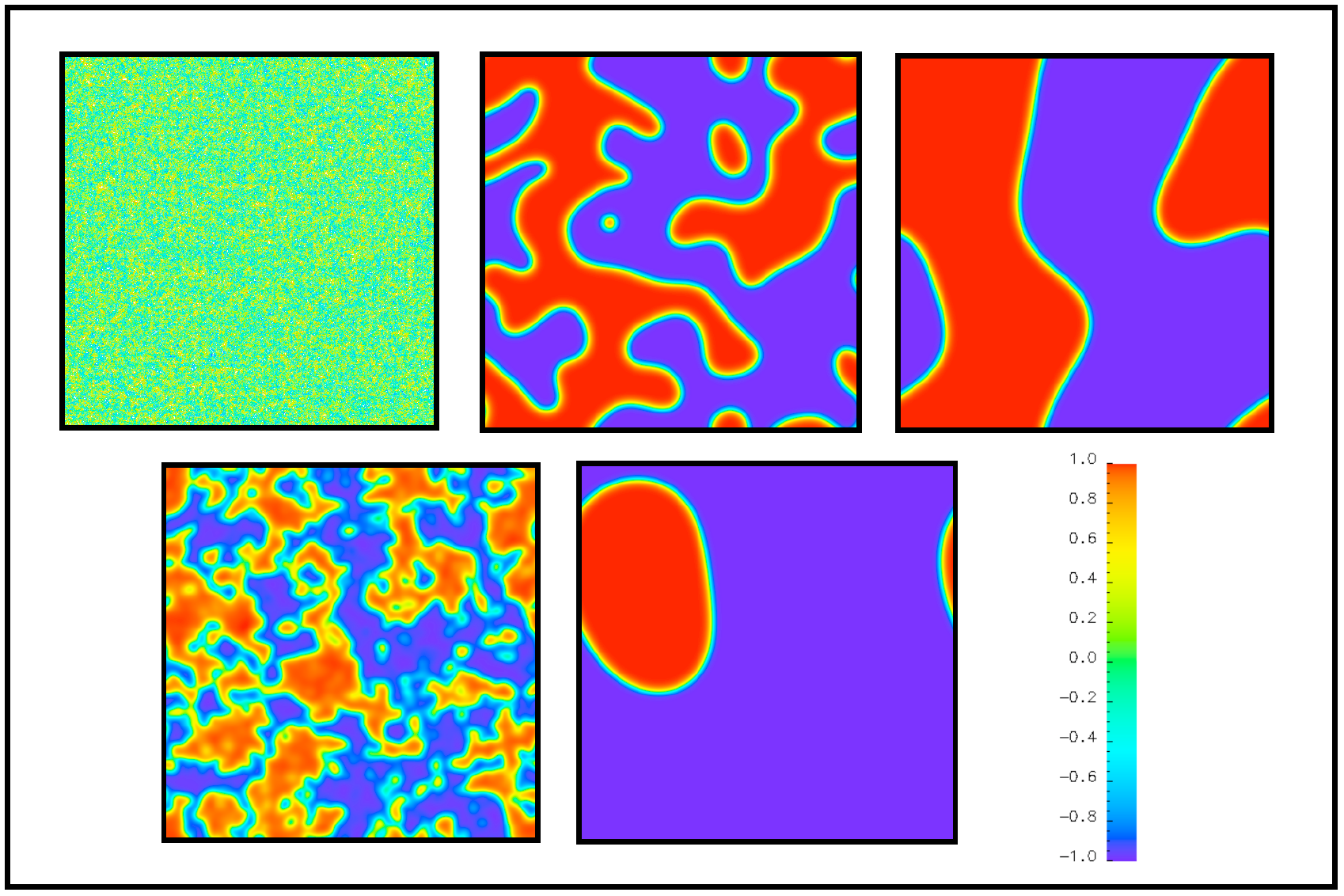}}
\caption{Evolution of $2d$ chiral domains. Red (+1 on the color bar) corresponds to the $L$-phase and blue (-1 on the color bar) corresponds to the $D$-phase. Time runs from left to right and top to bottom. Top left, the near-racemic initial conditions. Top mid and top right, evolution of the two percolating chiral domains separated by a thin domain wall. Bottom left, environmental effects break the stability of the domain wall network. Bottom right, subsequent surface-tension driven evolution leads to a enantiomerically-pure world \cite{GTW}. } \label{fig:2Dchiral}
\end{figure}

Extending this network to include spatio-temporal diffusion--and thus departing from the well-mixed limit described by ODEs--starts by substituting  $d/dt \rightarrow \partial / \partial t - k \nabla^2$, where $k$ is the diffusion constant \cite{BM}. In this coarse-grained approach, the number of molecules per unit volume is large enough so that the concentrations vary smoothly in space and time. The spatiotemporal evolution of the network is obtained by solving the coupled system of nonlinear PDEs for arbitrary values of $n$. Clearly, as $n$ increases, solving and statistically analyzing the coupled system of equations in two and three spatial dimensions for various parameters becomes highly CPU intensive. Following Soai \cite{Soai} where dimers were shown to be efficient catalysts (see also ref. \cite{BM}), GTW focused on the truncated system for $N = 2$ within the adiabatic approximation since, as with spatially-independent reaction networks, it has similar qualitative behavior to networks with longer ($N > 2$) polymer chains \cite{GW}. The system then reduces to two coupled PDEs for the concentrations $[L_1]$ and $[D_1]$, being thus more amenable to a detailed statistical study while maintaining the key qualitative features of a larger reaction network. Typical results are shown in the top three panels of Figure \ref{fig:2Dchiral}, where the two-phase system evolving from near-racemic conditions gets ``stuck'' due to the presence of both chiral phases.

The situation changes dramatically when noise is added to the system, as shown by GTW \cite{GTW}. This was motivated by experimental demonstrations that stirring can bias chirality \cite{Kondepudi, Viedma}, and numerical studies for a similar $N=2$ systems that explored how fluid turbulence can speed up chiral evolution \cite{BM}. Of course, the term ``punctuated'' makes reference to the ``punctuated evolution'' theory of Eldredge and Gould \cite{Eldredge}, where random, intense phenomena of widespread environmental impact resets the evolutionary soup, so to speak, followed by longer periods of stasis.
The role of the noise is to simulate random but substantial environmental perturbations that can affect the evolution of the biomolecular reaction network. In particular, the noise can, in principle, flip the direction of the chiral bias.  This is achieved in the simplest possible way via a generalized spatiotemporal Langevin equation \cite{GT}. The dynamical equations are written as:

\begin{eqnarray} \label{spatialeqns}
l_0^{-1}\left( \frac{\partial{\cal S}}{\partial t} - k \nabla^2 {\cal S} \right) &=&  1 - {\cal S}^2 + w(t, \textbf{x}),   \\ \nonumber
l_0^{-1}\left( \frac{\partial{\cal A}}{\partial t} - k \nabla^2 {\cal A} \right) &=& {\cal S}{\cal A} \left( \frac{2f}{{\cal S}^2 + {\cal A}^2} -1 \right)  +  w(t, \textbf{x}),   \\ \nonumber
\end{eqnarray}
where $l_0 \equiv (2k_SQ)^{1/2}$, and $w(\textbf{x},t)$ is a dimensionless Gaussian white noise with two-point correlation function $\langle w(\textbf{x}',t')w(\textbf{x},t) \rangle = a^2\delta(t'-t) \delta(\textbf{x}' -\textbf{x})$, where $a^2$ is a measure of the environmental influence's strength. An Ising phase diagram can be constructed showing that $\langle {\cal A} \rangle \rightarrow 0$ for $a > a_c$: chiral symmetry is restored \cite{GT}. The value of $a_c$ has been obtained numerically in two ($a_c^2 = 1.15(k/l_0^2)$cm$^2$s) and three ($a_c^2 = 0.65(k^3/l_0^5)^{1/2}$cm$^3$s) dimensions \cite{GT}. Dimensionless time, $t_0 = l_0 t $, and space, $x_0 = x(l_0/k)^{1/2}$, variables were introduced. For diffusion in water ($k = 10^{-9}$m$^2$s$^{-1}$) and nominal values $k_S = 10^{-25}$cm$^3$s$^{-1}$ and $Q = 10^{15}$ cm$^{-3}$s$^{-1}$, we obtain $l_0=\sqrt{2} \times 10^{-5}$s$^{-1}$, simulating a $2d$ ($3d$) shallow (deep) pool with linear dimensions of $l \sim 200$ ($50$) cm. For the purpose of illustration, explicit results quoted below were computed using these values. Details of the numerical implementation can be obtained in \cite{GTW}.

\begin{figure}
\centerline{\includegraphics[width=4in,height=3in]{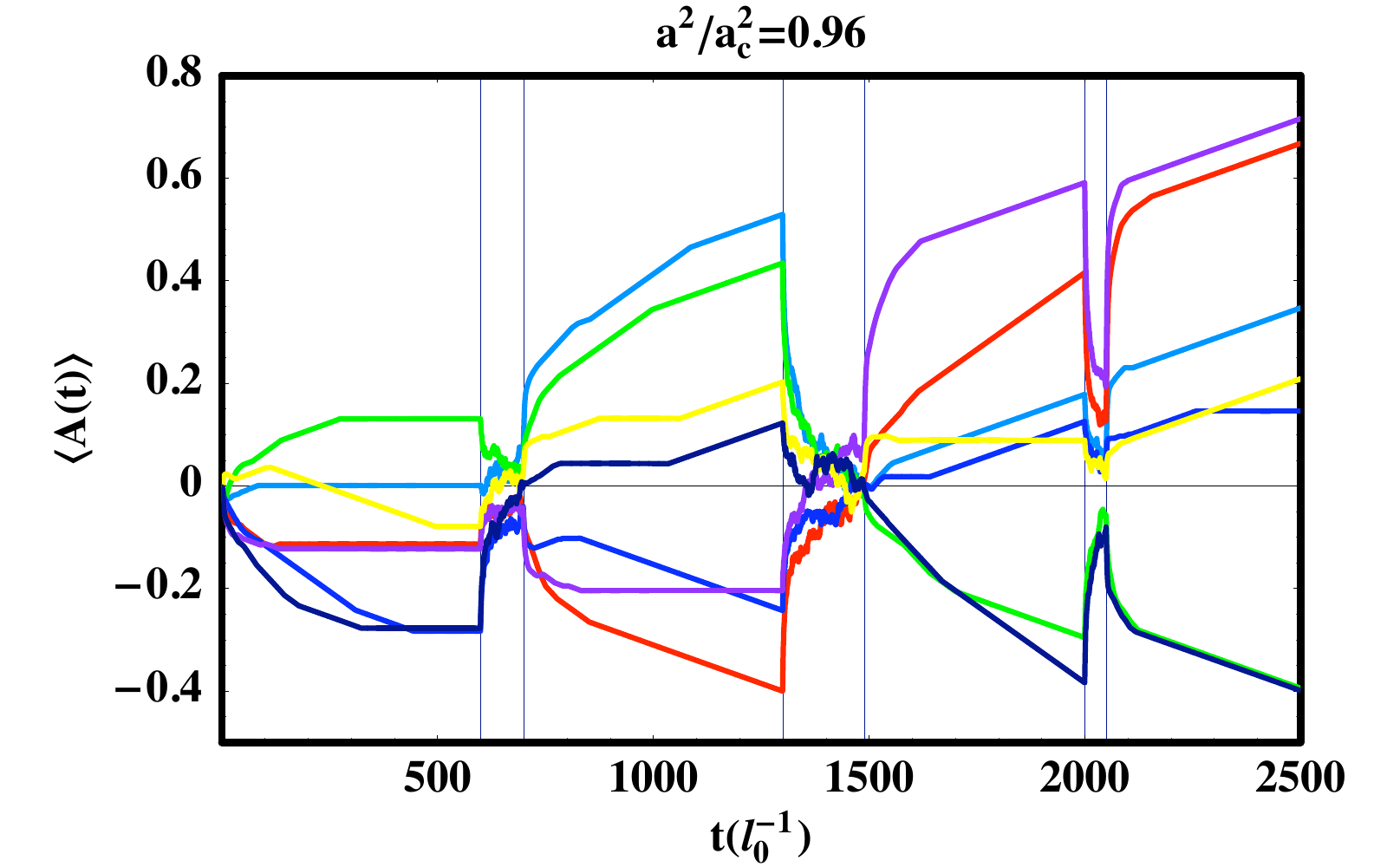}}
\caption{Punctuated Chirality \cite{GTW}. Impact of environmental effects of varying duration and fixed magnitude ($a^2/a_c^2 = 0.96$) on the evolution of prebiotic chirality in $2d$. Short events (last from left), which have little to no effect, should be contrasted with longer ones, which can drive the chirality towards purity and/or reverse its trend. (See, {\em e.g.} the green line.)} \label{fig:3event}
\end{figure}

Results of a large statistical sample of $100$ $2d$ runs that led to initial domain coexistence, that is, $d \langle {\cal A} \rangle /dt \approx 0$ show that near the critical region  $a^2 \geq 0.96 a_c^2$, all but the shortest events ($t \leq 50 l_0^{-1} \approx 1.5$ months, for the nominal value of $l_0=\sqrt{2} \times 10^{-5}$s$^{-1}$ mentioned previously) lead to statistically significant chiral biasing. Results in $3d$ are qualitatively very similar, although due to heavy CPU demand we limited the analysis to 50 short runs. Essentially, large environmental disturbances modeled here as Gaussian noise with a certain amplitude and duration can not only drive a near-racemic system toward homochirality but also can reverse the chirality of homochiral solutions, effectively erasing any previous chiral signature. 

As argued in GTW, the results suggest that, in the one extreme the early Earth may have played host to numerous abiogenetic events, only one of which ultimately led to the Last Universal Common Ancestor through the usual processes of Darwinian evolution. This is consistent with work indicating the widespread diversity and impact of extinction events, including the possibility of life emerging more than once \cite{Wilde}.  In the other extreme, one may consider, at the very least, that biological precursors certainly interacted with the primordial environment and may have had their chirality reset multiple times before homochiral life first evolved. In this case, our results show that separate domains of molecular assemblies with randomly set chirality may have reacted in different ways to environmental disturbances. A final, Earth-wide homochiral prebiotic chemistry would have been the result of multiple interactions between neighboring chiral domains under mechanisms described elsewhere \cite{G, GW, BM}. If punctuated chirality prevails, it implies that homochirality in different planetary platforms conducive to life or, at least, to stereochemistry is a random process, and thus localized. This means that a statistically large sample of extraterrrestrial stereochemistry would, on average, show no chiral bias. The Universe, taken as a whole, would be racemic.

\subsection{Chirality in Star-Forming Regions}

Laboratory investigations have shown that enantiomeric excesses can be produced by asymmetric photolysis or synthesis \cite{Griesbeck, Meierhenrich05}.
On the other hand, active star-forming regions can generate circularly polarized light (CPL) \cite{Bailey, CPL2, Lucas05}. In this case, and if the process is efficient, the stereochemistry of all worlds (meaning planets and moons) within this region should have a clear prevalence of one chiral bias over the other. The discovery of an enantiomeric excess of chiral organic compounds in the Murchison meteorite has supported this view for our solar system \cite{Cronin98,Glavin}. Furthermore, examination of isotopic distributions (in $^{15}$N/$^{13}$N) and of $\alpha$-branched amino acids of extraterrestrial origin has eliminated the possibility of contamination by Earth's biosphere, with measurements showing an excess of L-alanine of 50\% and of 30\% for L-glutamic acid \cite{Engel97}. A plausible explanation for such abiotic enantiomeric excess is that the star-forming region that originated the solar system was subjected to CPL, such as synchrotron radiation from a neutron star, although neutron stars don't appear to be a significant source of CPL in the visible and UV range \cite{Bailey}. If CPL-induced chiral bias is indeed a viable mechanism, the same chiral bias should be prevalent throughout the solar system but not necessarily throughout the galaxy. For example, finding an excess in D-chiral compounds in Mars \cite{Kminek,Glavin2020} or elsewhere in the solar system would contradict this scenario. (It would also contradict the possibility of a dominant chirality throughout the Universe, as we discuss in the next subsection.)

Chiral biasing through CPL depends on several unknowns such as the nature of the UV source, its distance from the target planet or moon, and the duration of effective radiative emission, making its viability harder to estimate  \cite{CPL2,KN85}. Furthermore, photolysis of amino acids requires UV radiation, which cannot be directly observed due to dust obscuration. Fukue {\it et al.} have investigated the range of CPL in the Orion nebula star forming region \cite{Fukue}. Although they reported a high circular polarization region with considerable spatially extension ($\sim 0.4$ pc; for comparison, the distance between the Sun and Proxima Centauri is $1.3$ pc) around the massive star-forming region known as the BN/KL nebula, other regions, including the linearly polarized Orion bar, show no significant circular polarization. Most of the low-mass young stars did not show detectable extended structure in either linear or circular polarization, in contrast to the BN/KL nebula. So, although CPL is a potential mechanism for generating an early bias toward enantiomeric excess, the only way to confirm its viability is by probing several worlds within the same original star-forming region. Certainly, finding meteorites with the same enantiomeric excess as on Earth strengthens this hypothesis \cite{Cronin98,Glavin}. However, the situation remains ambivalent. For example, Glavin and Dworkin didn't detect L-isovaline excess for the pristine Antarctic CR2 meteorites Elephant Moraine 92042 and Queen Alexandra Range 99177, whereas they did detect large levorotatory enantiomeric excesses in the CM meteorite Murchison and the CI meteorite Orgueil \cite{Glavin}. They suggest that this excess was produced by the amplification of a small initial excess by an aqueous alteration phase. 

Another obstacle to CPL-induced enantiomeric excess is its resilience against environmental disturbances as discussed above. A clear signature of a CPL-induced enantiomeric excess is that it must be correlated with a planet's formation era, as the meteoritic evidence suggests. If punctuated chirality is viable, it becomes difficult to explain the long-term resilience of a CPL-induced excess during, in the case of our planet, a window of a few hundred-million years, when the terrestrial environmental was certainly prone to large-scale perturbations due to heavy bombardment and active volcanism. Convincing evidence that CPL is the preferred mechanism for generating an enantiomeric excess would require a systematic search in most of our solar system. If, indeed, the same handedness prevails across our solar system, then CPL becomes a viable possibility. However, such evidence shouldn't be considered as the prevalent mechanism operating across the galaxy, unless parity-violation in the weak interactions can indeed be ruled out. We examine this mechanism next.

\subsection{Chirality from Parity Violating Interactions}

\noindent The discovery of parity violation in the weak interactions, and the subsequent formulation of electroweak theory in the Standard Model of particle physics suggests the possibility that parity-violating forces could affect quantum chemical calculations, including perturbative computations of parity violating potentials which, in turn, could play a role in biasing chirality at the molecular level \cite{WNC0,Salam91,KN85}. This would be an impressive connection between dynamics at the smallest level of elementary particle physics and the mechanisms that drive the homochirality of living systems, being, of course, of great aesthetic appeal. If this were the mechanism for biasing biological homochirality, prebiotic stereochemistry across the Universe would have to be the same, unifying the physics of the very small with cosmic stereochemistry and, possibly, life on Earth with life everywhere else in the Universe (if any).

Although initial results were small to the point of being negligible, more detailed calculations led to parity-violating energy differences of $\sim 10^{-11}$ Jmol$^{-1}$ (approximately 100 aeV) in enantiomers of chiral molecules \cite{WNC4,Quack08}. Clearly, even with this increase in the effect, a very efficient amplification mechanism is needed in order to affect molecular interactions of interest for biochemistry.

Kondepudi and Nelson \cite{KN85} constructed an ODE model where a small enantiomeric excess can be amplified by parity violation and is also subjected to CPL fluctuations. Unfortunately, their initially optimistic results were put into question when a more detailed spatiotemporal analysis was developed \cite{G}. This work used a similar parameterization as ref. \cite{KN85} to express the
bias due to parity violation in the weak nuclear interactions, which has been
estimated to be $g\equiv E_f/k_BT \sim 10^{-17-18}$ at room temperature
\cite{Salam91,WNC3,WNC4}. In ref. \cite{G} it was shown that using the well-mixed limit when discussing the dynamics greatly enhances the time scales where the chiral excess amplification can be effective. Indeed, including spatial dependence of the concentrations implies that the dynamics of symmetry breaking can take two possible paths, depending on the relative volume occupied by each phase: in the first case, if both phases are above percolation threshold, the system evolves as the domain walls separating the two phases vie for dominance. This domain-wall dynamics is clearly quite different from that of simple ODEs for the concentrations of enantiomers, and models the early Earth environment more realistically. Indeed, setting as a realistic time-scale for chirality to be defined as 100 million years implies that the parity-violating biasing must satisfy $g\geq 7\times 10^{-6}$(100My/$t_g)^{1/2}$ \cite{G}, unfortunately more than ten orders of magnitude larger than the quantum chemistry computations. For $g\leq 10^{-6}$ it would take longer than the age of the Universe before biasing becomes active in moving the walls toward a single phase. An estimate of the velocity with which the walls propagate shows that for a wall to convert a distance of 1km in 100My, $g\geq 4\times 10^{-6}$. So, it not only takes too long for the chiral bias $g$ to take over the dynamics of the domain wall system, but, once it does take over, the walls move too slowly to sweep an effective area in a realistic time scale $\sim 100$My.

In the second case, a two-phase system with one phase in a metastable state (the one unfavored by the chiral bias), and the walls moving slowly, we can examine how the dynamics evolves through bubble nucleation typical of first-order phase transitions \cite{G}. Unfortunately, the situation is worse in this case. If we impose the same time-scale for bubble nucleation as above,  $\tau_{\rm nuc}\leq 100$My, a ``shallow'' cylindrical pool with volume $\sim \pi\times 10^{8}{\rm m}^3$, gives $\tau_{\rm nuc} \simeq 4\times 10^{-18}\exp[89/g^2a^2]$y. Choosing a fairly large value $a^2\simeq 0.5$ for the typical noise amplitude that induces thermal fluctuations (cf. Fig. \ref{fig:3event}), we obtain $g\geq 1.74$ an unrealistically large value. One possibility left open is whether external perturbations could accelerate the decay of the metastable state. These could involve a variety of external influences, from cataclysmic events to meteoritic impact, perhaps already with some enantiomeric excess, thus combining punctuated chirality with phase transition dynamics. 

Although we cannot discard future advances in the simulation of phase dynamics or other potential amplification mechanisms such as those just mentioned, it seems unlikely at this point that parity violation played a role in biasing the biological homochirality observed on Earth. A strong test of this hypothesis would be to have a statistically representative sample of stereochemistry over many stellar systems. If the chiral bias toward a specific enantiomeric excess for all these samples is the same as on Earth, then we may infer that indeed parity violation has found a way to influence chirality across the Universe. The only caveats, of course, are, one, whether punctuated chirality is effective at a planetary scale and, two, whether unknown alien biological processes can also be effective in reversing chirality. If so, either or both of these processes would erase any memory of a prior uniform prebiotic chiral excess with varying levels of effectiveness across different planetary platforms, and we would be back to chirality as being contingent to a planet's specific geophysical history or biological processes.

\section{Concluding Remarks: Observing Biological Chirality in the Universe}

As mentioned above, clarifying the abiotic mechanism or mechanisms responsible for biasing a specific enantiomeric excess on Earth or elsewhere will rely on data collected from other planetary platforms, in this solar system and others \cite{Avnir}. The fundamental premise, inspired by what we know from life on Earth, is that even if the initial chiral biasing mechanism is abiotic, biological processes will incorporate this excess and take it to the next level. In terrestrial proteins, enzymatic function is dependent on the folding of amino acid chains into highly ordered structures, which is not possible without a chiral excess. Furthermore, since there are no known abiotic sources of enantiomeric excesses of D-amino acids or L-sugars on Earth, finding such excesses in another solar system planet or moon would be a strong indication of extraterrestrial biological processes \cite{Glavin2020}. Such a find would also support the hypothesis of punctuated chirality operating to generate distinct enantiomeric excesses at individual planetary platforms.

Several {\it in situ} and sample return missions are now in progress or under development. NASA's Curiosity rover \cite{Creamer} and the Philae lander of the Rosetta spacecraft \cite{Meierhenrich2013} were both equipped with gas-chromatographic coupled to mass spectrometry instruments capable of detecting enantiomeric excesses. Although at present {\it in situ} measurements of complex organics with enantioselective chromatography (coupled to a variety of detectors) are challenging to current spaceflight instrumentation, returning samples to Earth may be the best avenue to methodically and reliably search for enantiomeric excesses in extraterrestrial soil. Recent successful sample collection by the Perseverance rover opens the possibility that, when returned and analyzed on Earth by the end of the current decade, will allow the identification of chiral organic compounds on the Martian soil. The ESA, China, Japan, and Russia also have plans for sample collection and return. 

Another avenue for the identification of extraterrestrial chiral compounds is through remote sensing, such as the recent discovery of propylene oxide in the Sagittarius B2 star-forming region using a radio telescope \cite{McGuire2016}. This was the first chiral molecule discovered in outer space. Finding a chiral molecule in a star-forming region raises the prospect that chiral materials are available during the formation of proto-planetary disks, thus possibly being incorporated in nascent planets and diverse orbiting bodies such as asteroids and comets. This may well have been the case for our solar system. 

Radio astronomy is the primary method for studying the complex molecular content of interstellar clouds, including spectral features corresponding to fine-structure transitions of atoms or pure rotational transitions of polar molecules. In principle, such methods at high-precision and full polarization can even detect enantiomeric excesses, although that was not the case for ref. \cite{McGuire2016}. With this resource, we should hope for exciting results in the coming years where target star-forming regions could, in principle, reveal chiral organics with noticeable enantiomeric excesses. 

Looking further ahead, a statically significant sample of diverse star-forming regions with chiral organics will be invaluable in the determination of the mechanism(s) driving the enantiomeric excess. If they all show the same skewness, a universal causation mechanism acting across the galaxy (and possibly the Universe) such as parity violation would be strongly favored. Otherwise, CPL acting on different star-forming clouds would be favored, generating the same chiral excesses for each parent star region but different skewness from region to region. Either way, we would have strong observational evidence that the molecules that jump-started the abiotic processes that led to biological homochirality on Earth are operational across the galaxy. If such initial excess gets amplified or reversed due to environmental effects remain an open question. In this case, a  large sample that tends to a neutral enantiomeric excess would support punctuated chirality or other unknown mechanisms with similar planetary-wide impact. Finally, looking much further ahead, and inspired by the recent successful launch of the James Webb Space Telescope, we could imagine a scenario where exoplanets with promising biosignatures have their transit spectra examined at high resolution for signs of chiral biomolecules in their atmosphere. If the same compounds found on Earth are identified, we would have strong evidence for biotic activity in the exoplanetary atmosphere, pointing toward universal biochemical properties. Whether such compounds would have similar or opposite chirality to their terrestrial counterparts would reveal the predominant abiotic mechanisms that drive homochirality here and elsewhere in the cosmos.

\vspace{0.2in}

\noindent{{\bf Acknowledgements}: The author wishes to thank the editors of the present volume for the invitation to contribute.}

\vspace{0.2in}

\noindent{{\bf Conflict of Interest Statement}: The author declares no conflict of interest.}


\vspace{0.2in}

\centerline{\bf REFERENCES}

\begin{thebibliography}{}


\bibitem[Avnir 2021]{Avnir}
Avnir, D.:2021,
\newblock {Critical review of chirality indicators of extraterrestrial life}.
\newblock {\em New Astron. Rev.} {\bf 92}, 101596.


%\bibitem[\protect\citeauthoryear{Avnir}{2021}]{Avnir}
%Avnir, D.:2021,
%\newblock {Critical review of chirality indicators of extraterrestrial life}.
%\newblock {\em New Astron. Rev.} {\bf 92}, 101596.


\bibitem[Bailey 1998]{Bailey}
Bailey, J.:1998,
\newblock {Circular polarization in star-formation regions: implications for biomolecular homochirality}.
\newblock {\em Science} {\bf 281}, 672--674.

\bibitem[Bailey 2001]{CPL2}
Bailey, J.:2001,
\newblock {Astronomical sources of circularly polarized light and the
origin of homochirality}.
\newblock {\em Orig. Life Evol. Biosph.} {\bf 31}, 167--183.

\bibitem[Bakasov {\em et al.} 1998]{WNC4}
Bakasov, A., Ha, T.-K., and Quack, M.:1998,
\newblock {Ab initio calculation of molecular energies including parity
violating interactions}.
\newblock {\em J. Chem. Phys.} {\bf 109}, 7263--7285.

\bibitem[Blackmond 2019]{Blackmond2019}
Blackmond, D.G..:2019,
\newblock{The Origin of Biological Homochirality.}
\newblock{\em Cold Spring Harbor perspectives in biology}{\bf 11}(3), 032540.

\bibitem[Blackmond 2004]{Blackmond04}
Blackmond, D. G.:2004,
\newblock {Asymmetric autocatalysis and its implications for the
origin of homochirality}.
\newblock {\em PNAS} {\bf 101}, 5732--5736.


\bibitem[Brandenburg and Multam\"aki 2004]{BM}
Brandenburg, A. \& Multam\"aki, T.:2004,
\newblock {How long can left and right handed life forms coexist?}.
\newblock {\em Int. J. Astrobiol.}, {\bf 3}, 209--219.

\bibitem[Cataldo {\em et. al.} 2005]{Cataldo}
Cataldo, F., Brucato, J. R., \& Keheyan, Y.:2005,
\newblock {Chirality in prebiotic molecules and the phenomenon of photo- and radioracemization}.
\newblock {\em J. Phys.: Conf. Series} {\bf 6}, 139--148.

\bibitem[Cline 2000]{Cline}
Cline, D.:2000,
\newblock{The Physical Origin of Homochirality in Life (AIP Conference Proceedings)}
\newblock{\em AIP Conference Proceedings, 329}.

\bibitem[Nadir, N. 2016]{Creamer}
Creamer, J.S., Mora, M.F., \& Willis, P.A.:2016,
\newblock {Enhanced Resolution of Chiral Amino Acids with Capillary Electrophoresis for Biosignature Detection in Extraterrestrial Samples}.
\newblock {\em Anal. Chem.},{\bf 89}(2), 1329-1337.


\bibitem[Eldredge and Gould 1972]{Eldredge}
Eldredge, N., Gould, S.J.:1972,
\newblock {in {\em Models in Paleobiology} (ed. Schopf, T. J. M.)}
\newblock {Ch. 5 (Freeman Cooper, San Francisco, 1972).}

\bibitem[Engel and Macko 1997]{Engel97}
Engel, M. H. and Macko, S. A.;1997,
\newblock {Isotopic evidence for extraterrestrial non-racemic
amino acids in the Murchison meteorite}.
\newblock {\em Nature} {\bf 389}, 265--268.

\bibitem[Finefield {\em et al.} 2012]{Finefield2012}
Finefield, J.M. {\it et al.}.:2012,
\newblock {Enantiomeric natural products: occurrence and biogenesis}.
\newblock {\em Angew. Chem. Int. Ed.} {\bf 51}, 4802--4836.

\bibitem[Frank 1953]{Frank53}
Frank, F.~C.:1953,
\newblock {On spontaneous asymmetric catalysis}.
\newblock {\em Biochim. Biophys. Acta}, {\bf 11}, 459--463.

\bibitem[Fukue {\em et al.} 2010]{Fukue}
Fukue, T., et al.:2010,
\newblock {Extended High Circular Polarization in the Orion Massive Star Forming Region: 
Implications for the Origin of Homochirality in the Solar System}.
\newblock {\em Orig. Life Evol. Biosph.} {\bf 40}, 335--346.

\bibitem[Glavin and Dworkin 2009]{Glavin}
Glavin, D.P. \& Dworkin, J.P.:2005,
\newblock {Enrichment of the amino acid L-isovaline by aqueous alteration on CI and CM meteorite parent bodies}.
\newblock {\em Proc. Nat. Acad. Sci. } {\bf 106(14)}, 5487--5492.

\bibitem[Glavin {\em et al.} 2020]{Glavin2020}
Glavin, D.P. {\it et al.}:2020,
\newblock {The Search for Chiral Asymmetry as a Potential Biosignature in our Solar System}.
\newblock {\em Chem. Rev.} {\bf 120(11)}, 4660--4689.

\bibitem[Gleiser 2007]{G}
Gleiser, M.:2007,
\newblock {Asymmetric Spatiotemporal Evolution of Prebiotic Homochirality}.
\newblock {\em Orig. Life Evol. Biosph.}, {\bf 37}, 235--251.

\bibitem[Gleiser 2010]{Gleiser_book}
Gleiser, M.:2010,
\newblock{A Tear at the Edge of Creation: A radical new vision for life in an imperfect universe}
\newblock{\em Free Press, New York, NY}.

\bibitem[Gleiser and Thorarinson 2006]{GT}
Gleiser, M. \& Thorarinson, J.:2006,
\newblock {Prebiotic homochiralirty as a critical phenomenon}.
\newblock {\em Orig. Life Evol. Biosph.}, {\bf 36}, 501--505.

\bibitem[Gleiser, Thorarinson, Walker 2006]{GTW}
Gleiser, M., Thorarinson, J. \& Walker, S.I.:2008,
\newblock {Punctuated Chirality}.
\newblock {\em Orig. Life Evol. Biosph.}, {\bf 38}, 499--508.


\bibitem[Gleiser and Walker 2008]{GW}
Gleiser, M., \& Walker, S.I.:2008,
\newblock {An Extended Model for the Evolution of Prebiotic Homochirality: A Bottom-Up Approach to the Origin of Life}.
\newblock {\em Orig. Life Evol. Biosph.} {\bf 38}, 293--315.

\bibitem[Griesbeck and Meierhenrich 2002]{Griesbeck}
Griesbeck, A.G. \& Meierhenrich, U.J.:2002,
\newblock{Asymmetric Photochemistry and Photochirogenesis.}
\newblock{\em Angew. Chem. Int. Ed.}{\bf 41}, 3147--3154.

\bibitem[Hochberg 2021]{Hochberg21}
Hochberg, D. ed.:2021,
\newblock{Asymmetry in Biological Homochirality}
\newblock{\em Mdpi AG}


\bibitem[Janoschek 1991]{Janoschek}
Janoschek, R..:1991,
\newblock{Chirality from the Weak Bosons to the $\alpha$-helix}
\newblock{\em Springer Berlin Heildeberg, 1991.}

\bibitem[Kminek 2000]{Kminek}
Kminek, G., {\em et al} :2000,
\newblock {F. MOD: An organic detector for the future robotic exploration of Mars}.
\newblock {\em Planet Space Sci.}, {\bf 48}, 1087--1091.

\bibitem[Kondepudi and Nelson 1985]{KN85}
Kondepudi,~D.~K. and Nelson, G.~W.:1985,
\newblock {Weak Neutral Currents and the Origin of Biomolecular Chirality}.
\newblock {\em Nature}, {\bf 314}, 438--441.

\bibitem[Kondepudi {\em et. al.} 1990]{Kondepudi}
Kondepudi, D. K., Kaufman, R. J. \& Singh, N.:1990,
\newblock {Chiral Symmetry Breaking in Sodium Chlorate Crystallization}.
\newblock {\em Science}, {\bf 250}, 975--976.

\bibitem[Lazzeretti {\em et al.} 1999]{WNC3}
Lazzeretti, P., Zanasi R., and Faglioni F.:1999,
\newblock {Energetic stabilization of d-camphor via weak neutral currents}.
\newblock {\em Phys. Rev. E} {\bf 60}, 871--874.


\bibitem[Lucas {\em et al.} 2005]{Lucas05}
Lucas, P.~W. {\em et al.}:2005,
\newblock {UV Circular Polarization in Star Formation Regions:
The Origin of Homochirality?}.
\newblock {\em Orig. Life Evol. Biosph.}, {\bf 35}, 29--60.

\bibitem[McGuire {\em et al.} 2016]{McGuire2016}
McGuire, B.A., {\it et al.}2016,
\newblock {Discovery of the interstellar chiral molecule propylene oxide CH$_3$CHCH$_2$O}.
\newblock {\em Science} {\bf 352}(6292), 1449--1452.

\bibitem[Martin \& Russell 2007]{Martin2007}
Martin, W., and Russell, M. J. 2007,
\newblock {On the origin of biochemistry at an alkaline
hydrothermal vent}.
\newblock {\em Phil. Trans. R. Soc. B} {\bf 362}, 1887--1925.


\bibitem[Meierhenrich {\em et al.} 2005]{Meierhenrich05}
Meierhenrich, U.J., Mu\~noz, C., Schutte, W.A., Thiemann, W.H-P., Barbier, B., \& Brack, A.:2005,
\newblock{Precursors of biological cofactors from ultraviolet irradiation of circumstellar/interstellar ice analogs.}
\newblock{\em Chem. Eur. J.}{\bf 11}, 4895--4900.

\bibitem[Meierhenrich {\em et al.} 2013]{Meierhenrich2013}
Meierhenrich, U.J., {\it et al}.:2013,
\newblock{Evaluating the robustness of the enantioselective stationary phases on the Rosetta mission against space vacuum vaporization}
\newblock{\em Adv. Spac. Res.}{\bf 52}(12), 2080--2084.



\bibitem[Pasteur 1848]{Pasteur}
Pasteur, L.:1848,
\newblock {Recherches sur les relations qui peuvent exister entre la forme crystalline et la composition chimique, et le sens de la polarization rotatoire}.
\newblock {\em Ann. Chim. Phys.}, {\bf 24}, 442--459.

\bibitem[Pizzarello and Cronin 1998]{Cronin98}
Pizzarello, S. and Cronin, J. R.:1998,
\newblock {Alanine enantiomers in the Murchison meteorite}.
\newblock {\em Nature} {\bf 394}, 236.

\bibitem[Soai {\em et. al.} 1995]{Soai}
Soai, K., Shibata, T., Choji, K. \& Morioka, H.:1995,
\newblock {Asymmetric autocatalysis and amplification of enantiometric
excess of a chiral molecule}.
\newblock {\em Nature} {\bf 378}, 767--768.


\bibitem[Quack, Stohner, Willeke 2008]{Quack08}
Quack, M., Stohner, J., \& Willeke, M.:2008,
\newblock {High-Resolution Spectroscopic Studies and Theory of Parity Violation in Chiral Molecules}.
\newblock {\em Ann. Rev. Phys. Chem}, {\bf 59}, 741--769.

\bibitem[Salam 1991]{Salam91}
Salam, A.:1991,
\newblock {The Role of Chirality in the Origin of life}.
\newblock {\em J. Mol. Evol.}, {\bf 33}, 105--113.

\bibitem[Sandars 2003]{Sandars03}
Sandars, P.~G.~H.:2003,
\newblock {A Toy Model for the Generation of Homochirality
During Polymerization}.
\newblock {\em Orig. Life Evol. Biosph.}, {\bf 33}, 575--587.

\bibitem[Viedma 2005]{Viedma}
Viedma, C.:2005,
\newblock {Chiral symmetry breaking during crystallization: complete chiral purity induced by nonlinear autocatalysis and recycling}.
\newblock {\em Phys. Rev. Lett.} {\bf 94},065504.


\bibitem[Wagniere 2007]{Wagniere}
Wagni\`ere, G.H.:2007,
\newblock {On chirality and the universal asymmetry : reflections on image and mirror image}.
\newblock {\em VHCA with Wiley-VCH, Z\"urich, 2007}.

\bibitem[Wattis and Coveney 2005]{WC}
Wattis, J.~A. and Coveney, P.~V.:2005,
\newblock {Symmetry-Breaking in Chiral Polymerization}.
\newblock {\em Orig. Life Evol. Biosph.}, {\bf 35}, 243--273.

\bibitem[Wilde {\em et al.} 2001]{Wilde}
Wilde, S.A., Valley, J.W., Peck, W.H., \& Graham, C.M.:2001,
\newblock {Evidence from detrital zircons for the existence of continental crust and oceans on the Earth 4.4?Gyr ago}.
\newblock {\em Nature}, {\bf 409}, 175--178.


\bibitem[Yamagata 1966]{WNC0}
Yamagata, Y.:1966,
\newblock {A hypothesis for the asymmetric appearance of biomolecules on earth}
\newblock {\em J. Theoret. Biol.} {\bf 11}, 495--498.



\end{thebibliography}
\end{document}